# Landau Level Quantization and Possible Superconducting Instabilities in Highly Oriented Pyrolitic Graphite

Y. Kopelevich [1], V. V. Lemanov [2], S. Moehlecke [1], and J. H. S. Torres [1]

[1] Instituto de Física "Gleb Wataghin", Universidade Estadual de Campinas, Unicamp 13083-970, Campinas, São Paulo, Brasil

[2] A. F. Ioffe Physico-Technical Institute, Russian Academy of Sciences, 194021 St. Petersburg, Russia

Abstract

Measurements of the basal-plane resistivity $\rho_a(T,H)$ performed on highly oriented pyrolitic graphite, with magnetic field H ∥ c-axis in the temperature interval 2 - 300 K and fields up to 8 T, provide evidence for the occurrence of both field-induced and zero-field superconducting instabilities. Additionally, magnetization M(T,H) measurements suggest the occurrence of Fermi surface instabilities which compete with the superconducting correlations.

The magnetic field - temperature (H-T) phase diagram of conventional type-II superconductors is well known. In the Meissner state, the surface currents screen the applied magnetic field. Above the lower critical field $H_{c1}(T)$, the field penetrates the superconductor in the form of a lattice of vortices (Abrikosov lattice). Superconductivity persists up to the upper critical field $H_{c2}(T)$, described by the Abrikosov-Gor'kov theory [1, 2]. On the other hand, it has been proposed [3] that the superconducting state can appear (or reappear) under application to an electron system of high enough magnetic field, such that the Landau quantization of the energy spectrum is important. In particular, when all electrons are in the lowest Landau level, the superconducting transition temperature $T_c(H)$ is expected to increase with field increasing, opposite to the $T_c(H)$ dependence in the classical low-field-limit [3]. However, superconductivity in the quantum regime has not been identified in experiments so far, remaining the subject of theoretical investigations only.

In the present work, we report the results of basal-plane resistivity $\rho_a(T,H)$ measurements performed on highly oriented pyrolitic graphite (HOPG), which provide evidence for the occurrence of superconducting correlations in both quantum and classical limits. Besides, magnetization measurements $M(T,H)$ suggest an interplay between superconducting and other Fermi surface instabilities, possibly spin-density-wave (SDW) or charge-density-wave (CDW) type.

The HOPG sample was obtained from the Research Institute "GRAPHITE" (Moscow). X-ray diffraction ($\Theta$ - $2\Theta$) measurements give the crystal lattice parameters a = 2.48 Å and c = 6.71 Å. The high degree of crystallites orientation along the hexagonal c-axis was confirmed from x-ray rocking curves (FWHM = 1.4º). The geometrical sample density was $2.26 \pm 0.01$ g/cm$^3$. A cylindrical specimen with diameter of 5.2 mm, and thickness 3.14 mm, and a parallelepiped 4.9 x 4.3 x 2.5 mm$^3$, both made from the same piece of HOPG, were used for magnetization and transport measurements, respectively. The c-axis was along the smallest size of the sample. The studies were performed for H ∥ c-axis. M(T,H) dc magnetization was measured in fields up to 5 T and temperatures between 2 and 300 K by means of SQUID magnetometer MPMS5 (Qunatum Design). Low-frequency (f = 1 Hz) standard four-probe resistance measurements were performed in fields up to 8 T, in the same temperature interval, with PPMS (Physical Properties Measurement System, Quantum Design).

Low-temperature portions of the basal-plane resistivity $\rho_a(T)$ measured for magnetic fields H ≤ 0.08 T are shown in Fig. 1. As can be seen from Fig. 1, $\rho_a(T)$ has a well defined maximum at a temperature $T_{max}(H)$ (as defined in the inset of Fig. 1) which is decreasing function of field. Thus, $T_{max}(H)$ separates a high-temperature semiconducting-like behavior ($\rho_a$ increases with temperature decreasing) from a low-temperature metallic-like behavior ($\rho_a$ decreases with temperature decreasing). In the field interval 0.08 T < H < 2.6 T, the maximum in $\rho_a(T)$ does not occur (Fig. 2). With a further increase in the field, the maximum in $\rho_a(T)$ can be observed again for certain H (Fig. 3). A non-monotonous behavior of $T_{max}$ vs. H, and the competition between the metallic-like and semiconducting-like behavior can be seen in Fig. 3. At H ≥ 3.9 T, $T_{max}(H)$ occurs at all measuring fields. Temperature dependences of $\rho_a(T)/\rho_a(T_{max})$ vs. T for several fields in the interval 4 T ≤ H ≤ 8 T are shown in Fig. 4. In this high-field regime, $T_{max}$ increases with H increasing.

Temperature dependences of normalized magnetization M(T)/|M(2K)| at various applied fields are presented in Figs. 5 and 6. The inset to Fig. 6 shows M(T) measured for H = 4, 4.5 and 5T demonstrating that the absolute value of diamagnetic magnetization M(T,H) increases with field increasing and temperature decreasing, in

agreement with previous reports [4]. The novel feature is the occurrence of a minimum in M(T) (Fig. 5). At low applied fields, H < 0.3 T, the minimum in M(T) takes place at nearly field-independent temperature $T_{min}$ = 32 - 35 K. At H > 0.3 T, $T_{min}$(H) is a non-monotonous function of the field. For fields H > 3 T, |M(T)| monotonously increases with temperature decreasing.

Fig. 7 presents magnetization M(H) and susceptibility $\chi$ = dM/dH vs. H at T = 2 K. As seen in Fig. 7, $\chi$(H) exhibits pronounced oscillatory behavior in the field interval 1 < H < 3.5 T due to de Haas-van Alphen effect coupled to the Landau level quantization. The reduction of de Haas-van Alphen oscillations at H > 3.5 T indicates that carries occupy only lowest Landau levels at higher fields.

All the experimental results are summarized in Fig. 8, where points 1 and 2 correspond to $H(T_{max})$ and $H(T_{min})$, respectively. The inset in Fig. 8 depicts a high-field portion of the $H(T_{max})$, plotted in a linear scale.

The resistivity drop below $T_{max}$(H) can be understood assuming the occurrence of Fermi surface instabilities at $T_{max}$(H) with respect to the Cooper pairs formation.

Actually, the rapid increase of $T_{max}$ with field increasing (H > 3.9 T, see the inset in Fig. 8) resembles very much that of the superconducting transition temperature $T_c$(H) in the quantum limit (H > $H_{QL}$), where carriers are in the lowest Landau level [3]. The $T_c$(H) given by [3]

$$T_c(H > H_{QL}) = 1.14\Omega \exp[-2\pi l^2/N_1(0)V], \qquad (1)$$

results from the increase in a 1D density of states $N_1(0)$ at the Fermi level, where $2\pi l^2/N_1(0) \sim 1/H^2$, $l = (\hbar c/eH)^{1/2}$, V is the BCS attractive interaction, and $\Omega$ is the energy cutoff on V. With a further increase in field, a saturation in $T_c$(H) followed by a reduction of $T_c(H)$, is expected [3]. Thus, the saturation in $T_{max}(H)$ occurring for $H > 6$ T, see Fig. 8, is consistent with the predicted $T_c(H)$ behavior. One of the reasons for the suppression of $T_c(H)$ is the Zeeman splitting, leading to a destruction of the spin-singlet superconductivity above a spin-depopulation field $H_d > H_{QL}$. It is important to note that a relatively small effective g-factor of graphite, $g^* = (m^*/m_0)g \sim 0.1$, ensures a substantial field interval above $H_{QL}$ where both spin-up and spin-down states should be occupied [3] (here $m^*_e/m_0 = 0.058(9)$, $m^*_h/m_0 = 0.04$ are effective masses of the majority electrons and majority holes, divided by free-electron mass [5, 6], and $g \approx 2$ [7]). For $H < H_{QL}$, the theory predicts an oscillatory behavior of $T_c(H)$ [3, 8-10], which is also in excellent agreement with the non-monotonous $T_{max}$ vs. H behavior, found in the regime of pronounced Landau level oscillations (Fig. 3 and 8).

On the other hand, as emphasized in Ref. [3], the high-field superconductor can be a non-superconducting material in the classical low-field-limit. Assuming, however, that superconducting instabilities are responsible for the resistance drop at $T < T_{max}(H)$ for all studied fields, one tends to verify the relation [3]

$$H_{QL} \sim (E_F/T_{c0})^2 H_{c2}(0). \qquad (2)$$

Interestingly, the behavior $H(T_{max}) \sim (T_{max}(0) - T)^{0.5}$ (dotted line in Fig. 8) perfectly agrees with the upper critical field behavior of granular superconductors near $T_c$ [11, 12]

$$H_{c2}(T) \sim (T_{c0} - T)^\alpha, \qquad (3)$$

where $\alpha = 0.5$ is the characteristic exponent of inhomogeneous systems of nearly isolated superconducting grains, and $T_{c0}$ is the zero-field superconducting transition temperature. Taking the Fermi energy $E_F = 0.024$ eV [5], and considering $T_{c0} \sim 50$ K and $H_{c2}(0) \sim 0.1$ T, one calculates using Eq. (2) that $H_{QL} \sim 2.3$ T. This value is close to the experimental value H = 3.9 T above which $T_{max}$ monotonously increases with H. Supporting the occurrence of superconducting instabilities in the low-field-limit, the resistivity below ~ 50 K exhibits a strong field dependence (see Fig. 1), consistent with the field-induced suppression of superconducting correlations.

In the magnetization measurements, the lack of evidence for the Meissner effect should be noted, first of all. The absence of the Meissner effect at high fields is in agreement with the theory of superconductivity in the quantum limit [3, 10]. On the other hand, at low fields Meissner effect can be not seen due to the small size of superconducting regions (grains). The Meissner effect can also be masked by the proximity of $T_{max}(H)$ to $T_{min}(H)$, below which |M(T)| decreases. We stress that the non-monotonous behavior of $T_{min}$ vs. H (Figs. 5 and 8) excludes a trivial origin of the magnetic anomaly, such as arising, e. g., from paramagnetic impurities. In search for an explanation of the minimum in M(T), one should take into account that all contributions to the temperature-dependent magnetization of graphite come from carrier states situated in a vicinity of the Fermi level [5]. At the same time, Fig. 8 demonstrates that at small fields $H(T_{max})$ line terminates exactly at the $H(T_{min})$ boundary, and that $T_{min}(H)$ rapidly increases with field above ~ 2 T, where $T_{max}(H)$ reappears. Based on these observations, it is tempting to conclude that Fermi surface instabilities, competing with superconducting instabilities, are responsible for the magnetic anomaly. These can be either CDW or SDW, both enhanced at high fields due to increase in $N_{1n}(0)$, the 1D density of states for the n-th Landau level [3], which explains the $T_{min}(H)$ increase for H

> 2 T. One may further speculate that CDW or SDW states overcome the superconducting correlations at H > 0.08 T, while superconducting correlations are stronger in the quantum limit (H ≥ 3.9 T). The tendency to saturation in $\rho_a(T)$ at T < $T_{max}(H)$, in low fields (Fig. 1), as well as the "reentrant" ($d\rho_a/dT < 0$) behavior observed for high fields (Figs. 3 and 4) are also consistent with the competition between superconductivity and CDW or SDW. At the same time, such resistance behavior is characteristic of inhomogeneous (granular) superconductors (see, e. g., refs. [12 - 16]). Here, a further both experimental and theoretical work is needed.

Finally, we want to comment on the semiconducting-like high-temperature behavior of $\rho_a(T)$. The HOPG is a polycrystalline layered material with a random orientation of crystallites within the layers. Thus, one may assume that the $\rho_a(T)$ is governed by the inter-crystallite boundaries. However, our zero-field value of $\rho_a(300 K) \approx 45$ μΩ·cm nearly coincides with the single-crystal resistivity value [17]. Therefore, we conclude that the inter-crystallite boundary effect is negligible. On the other hand, the $\rho_a(T)$ may originate from a reduced overlap of π orbitals, leading to a reduced carrier mobility, and the dominant effect of carrier density (which decreases with temperature decreasing) on $\rho_a(T)$. Note also, that the decrease in the π-electron overlap would imply an increase in the density of states, responsible for the occurrence of superconducting correlations at high temperatures in our HOPG.

In conclusion, we demonstrated the experimental evidence for the magnetic-field-induced superconducting instabilities due to Landau level quantization and the occurrence of zero-field superconducting correlations at $T_{c0} \approx 50$ K in the highly oriented pyrographite.

This work was partially supported by FAPESP proc. No. 95/4721-4, proc. No. 98/14726-1, proc. No.99/00779-9, CNPq proc. No. 300862/85-7, proc. No. 301216/93-2, and CAPES proc. No. DS-44/97-0.

**REFERENCES**

[1] A. A. Abrikosov, Sov. Phys. JETP **5**, 1174 (1957).

[2] L. P. Gor'kov, Sov. Phys. JETP **9**, 1364 (1959).

[3] M. Rasolt, Z. Tešanovic, Rev. Mod. Phys. **64**, 709 (1992) and references therein.

[4] J. Heremans, C. H. Olk, and D. T. Morelli, Phys. Rev. B **49**, 15122 (1994).

[5] M. P. Sharma, L. G. Johnson, and J. W. McClure, Phys. Rev. B **9**, 2467 (1974).

[6] N. B. Brandt, A. S. Kotosonov, S. V. Kuvshinnikov, and M. V. Semenov, Sov. Phys. JETP **52**, 476 (1980).

[7] G. Wagoner, Phys. Rev. **118**, 647 (1960).

[8] A. H. MacDonald, H. Akera, and M. R. Norman, Phys. Rev. B **45**, 10147 (1992).

[9] M. R. Norman, H. Akera, and A. H. MacDonald, Physica C **196**, 43 (1992).

[10] H. Akera, A. H. MacDonald, and M. R. Norman, Physica C **184**, 337 (1993).

[11] G. Deutscher, O. Entin-Wohlman, and Y. Shapira, Phys. Rev. B **22**, 4264 (1980).

[12] B. I. Belevtsev, Sov. Phys. Usp. **33**, 36 (1990).

[13] R. Laibowitz, A. Broers, D. Stroud, and B. R. Patton, AIP Conf. Proc. No. 58, ed. by H. C. Wolfe (A. I. P. New York, 1980), p. 278.

[14] H. M. Jaeger, D. B. Haviland, A. M. Goldman, and B. G. Orr, Phys. Rev. B **34**, 4920 (1986).

[15] A. Gerber, T. Grenet, M. Cyrot, and J. Beille, Phys. Rev. Lett. **65**, 3201 (1990).

[16] N. V. Zaitseva, Y. Kopelevich, I. I. Kochina, V. V. Lemanov, and P. P. Syrnikov, Sov. Phys. Solid State **33**, 323 (1991).

[17] M. S. Dresselhaus and G. Dresselhaus, Adv. Phys. **30**, 139 (1981) and references therein.

**FIGURES**

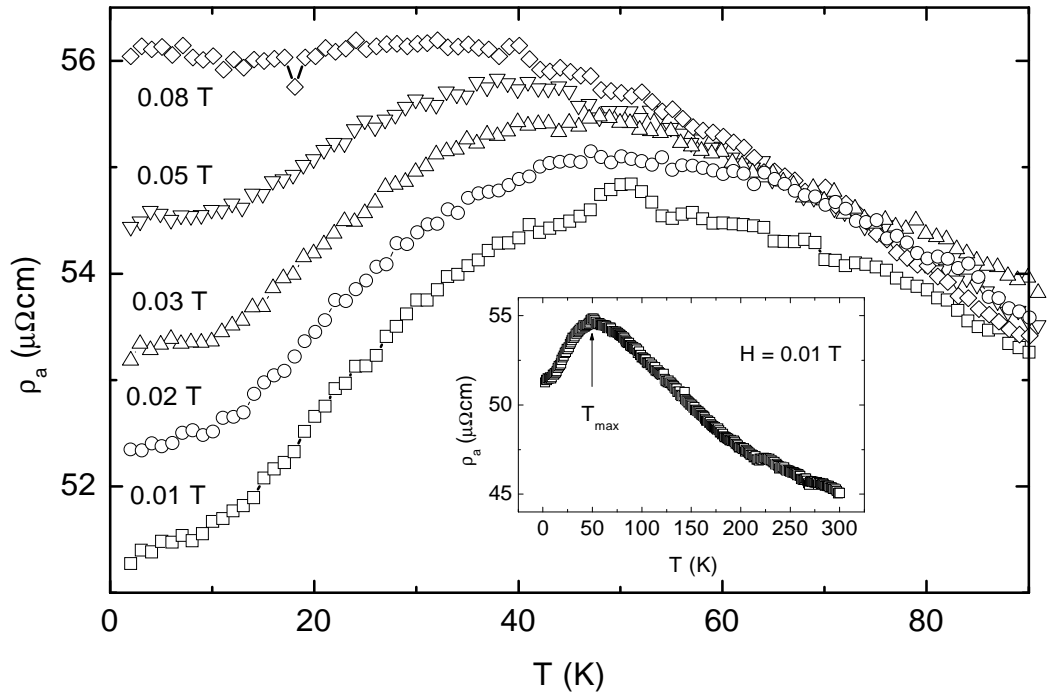

Fig. 1. Basal-plane resistivity $\rho_a(T)$ in the low-field-limit. Inset shows $\rho_a(T)$ obtained in the whole temperature interval under study for H = 0.01T.

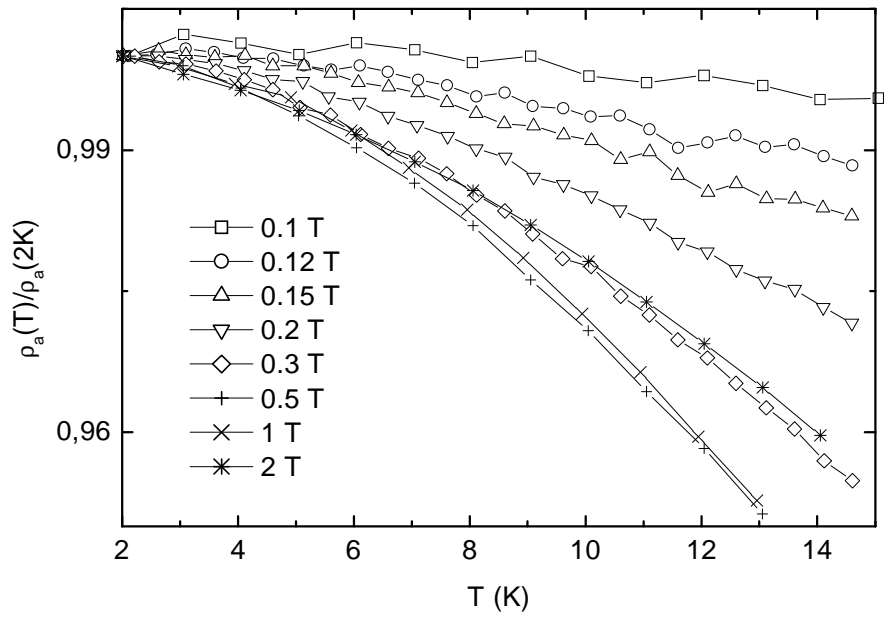

Fig. 2. Normalized resistivity $\rho_a(T)/\rho_a(2K)$ in the field interval where $T_{max}(H)$ does not occur.

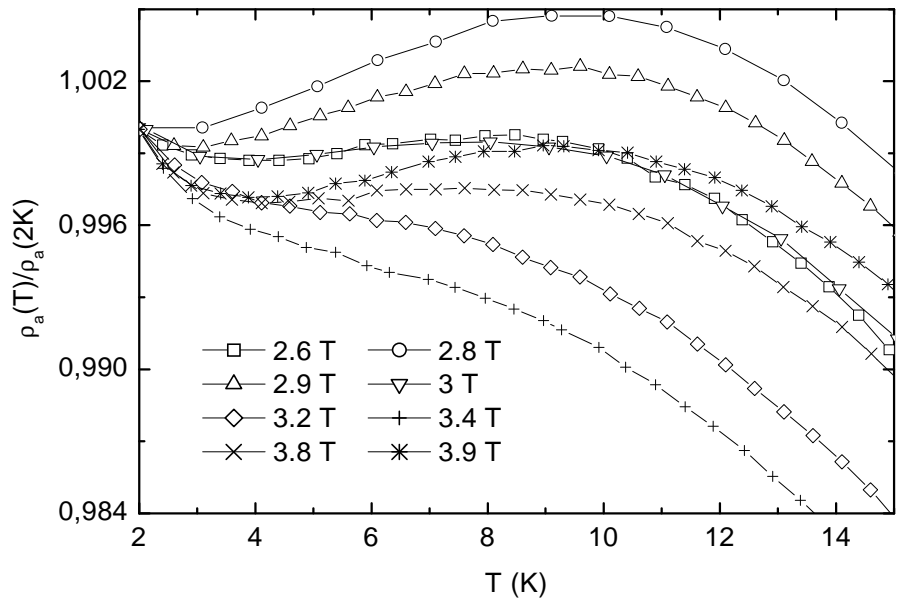

Fig. 3. Normalized resistivity $\rho_a(T)/\rho_a(2K)$ measured in the field interval where $T_{max}(H)$ reappears.

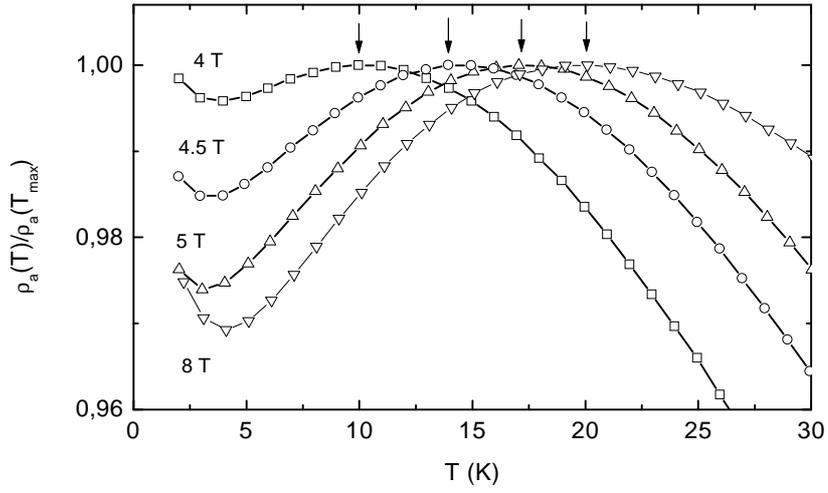

Fig. 4. Normalized resistivity $\rho_a(T)/\rho_a(T_{max})$ measured in the quantum limit for several fields. Arrows indicate $T_{max}(H)$.

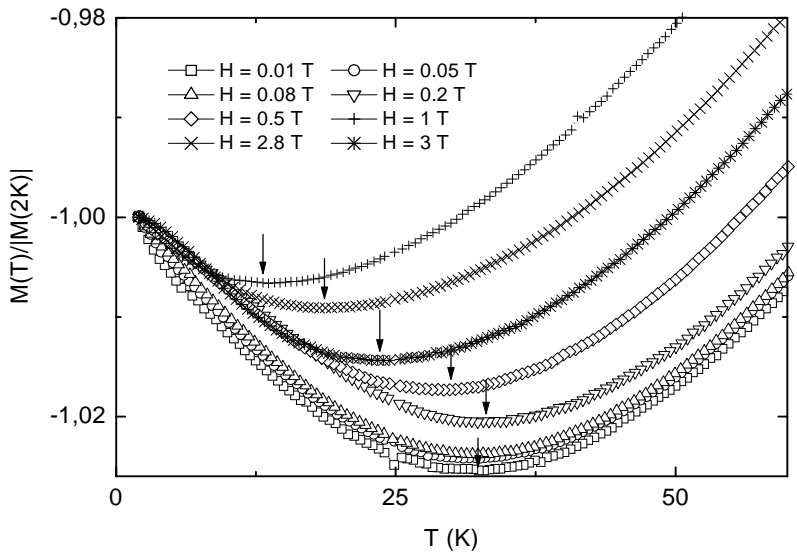

Fig. 5. Normalized magnetization $M(T)/|M(2K)|$ for various fields. Arrows indicate $T_{min}(H)$.

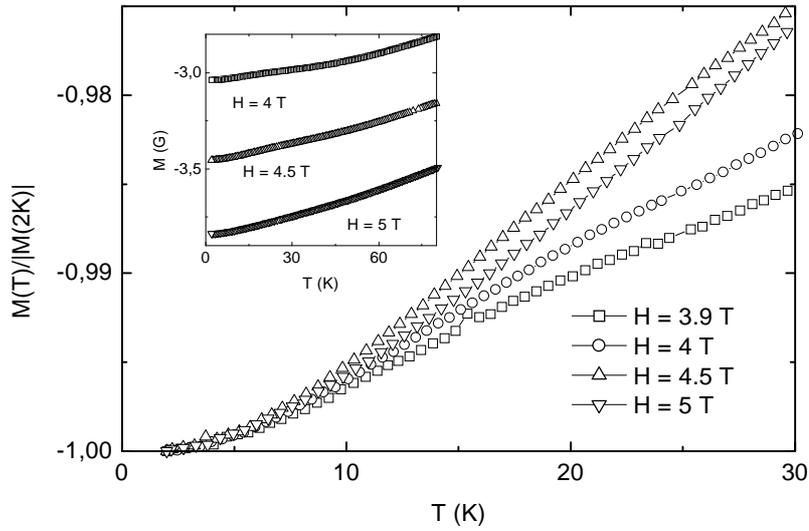

Fig. 6. Normalized magnetization M(T)/|M(2K)| in the quantum limit for several fields. Inset exemplifies temperature dependences of magnetization at H = 4 , 4.5 and 5 T.

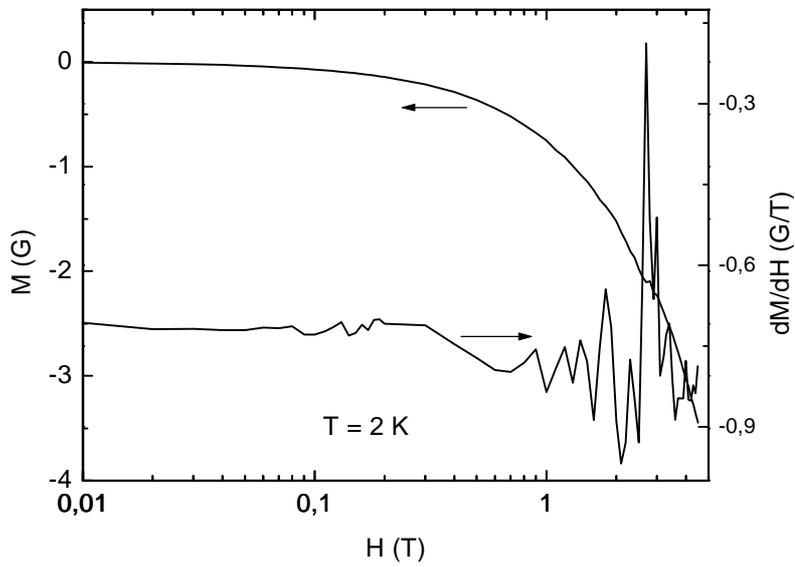

Fig. 7. Magnetization M and susceptibility $\chi$ = dM/dH vs. H obtained at T = 2 K. Susceptibility oscillations are due to de Haas-van Alphen effect.

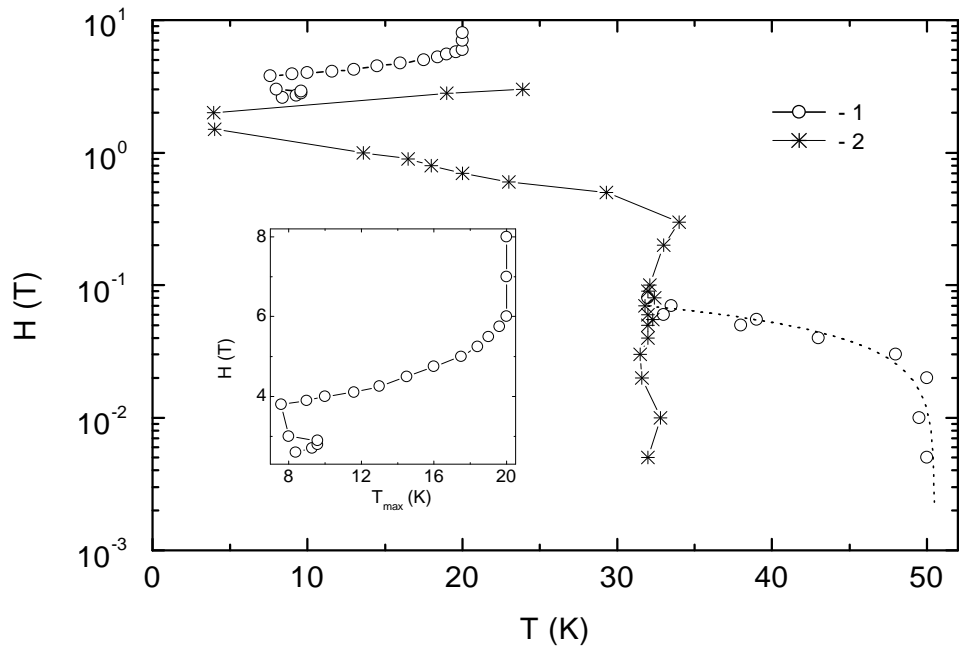

Fig. 8. Magnetic field - temperature diagram constructed from $\rho_a(T, H)$ and $M(T, H)$ data. 1 - $H(T_{max})$ obtained from $\rho_a(T, H)$, 2 - $H(T_{min})$ obtained from $M(T, H)$. The dotted line is the fit to the upper critical field boundary (see text) $H_{c2}(T) = A(1-T/T_{c0})^{0.5}$ with the fitting parameters $A = 0.115$ T and $T_{c0} = 50.5$ K. Inset presents a linear plot of H vs. $T_{max}$, measured at high fields.